\documentclass[12pt]{article}
\begin{document}
\begin{center}
\large\bf{The Renormalization Group\\
with Exact $\beta$-functions}
\end{center}
\vspace{2cm}
\begin{center}
V. Elias\\
D.G.C. McKeon\\
Department of Applied Mathematics\\
University of Western Ontario\\
London\\
CANADA\\
N6A 5B7
\end{center}
\vspace{1cm}
Tel: (519)661-2111, ext. 88789\hfill PACS No. 11.10Z\\
Fax: (519)661-3523\\
email: dgmckeo2@uwo.ca \\
velias@uwo.ca\vspace{4cm}\\

\section{Abstract}
The perturbative $\beta$-function is known exactly in a number of supersymmetric theories and in the 't Hooft renormalization scheme in the $\phi_4^4$ model. It is shown how this allows one to compute the effective action exactly for certain background field configurations and to relate bare and renormalized couplings. The relationship between the MS and SUSY subtraction schemes in $N = 1$ super Yang-Mills theory is discussed.

\section{Introduction}
In a number of instances, the perturbative renormalization group $\beta$-function is known exactly. 
In $N = 4$ supersymmetric Yang-Mills (SYM) theory, an $\beta$-function vanishes [1,2]. In $N = 2$ SYM theory, the $\beta$-function is exact at one-loop order when minimal subtraction (MS) is used [3,4]. In $N = 1$ SYM theory, the all-orders expression for the $\beta$-function can be determined either through instanton calculus [5] or by considering the multiplet structure of anomalies [6,7], though such an expression differs from the perburbative result derived using MS [8,9].
Although models such as the $\phi_4^4$ scalar theory and Yang-Mills (YM) theory have all-orders contributions to the $\beta$-function in the MS scheme, one can nevertheless perform in principle a finite renormalization at each order of perturbation theory so as to have contributions to the $\beta$-function vanish beyond two-loop order and to have anomalous dimensions as entirely one-loop effects [10,11].

In this paper we demonstrate how knowledge of the full $\beta$-function can be used to extract information about the effective action. In the first instance, the effective Lagrangian in a background $U(1)$ gauge field is determined for $N = 1$ and $N = 2$ SYM theories. In making this determination, we exploit the fact that the trace of the energy-momentum tensor $\theta_{\mu\nu}$ is proportional to the $\beta$-function [12-15]. In ref. [16] this proportionality is used in spinor and scalar QED to determine the $\beta$-function to two-loop order from the two-loop effective action computed in the presence of a self-dual background electromagnetic field. Our approach here is to employ a known $\beta$-function to determine the effective action in the presence of a background vector field that gives rise to $\theta_{\,\mu}^\mu$.

By having restricted the background field to being a vector field and not having included any contribution from background spinor fields, SUSY invariance is lost. This sort of effective action has been considered in (46, 47) where the Euler-Heisenberg
effective action in $N = 4$ SYM theory has been computed to one loop order.

By ``effective action'' we are referring to the sum of all one particle irreducible diagrams whose external legs are vector fields
corresponding to a constant field strength. This differs from the effective action considered in refs. [49, 50] where manifest supersymmetry is maintained by forming superfields out of composite fields and heuristicly defining the effective Lagrangian in a way that incorporates the Ward identities and the anomaly in the divergence of the supercurrent.

We next show how knowledge of the exact $\beta$-function can relate bare and renormalized couplings in closed form, using the approach developed in of ref. [17].

The summation of the logarithmic corrections to the effective potential in a simple $\phi_4^4$ scalar model is then examined in the context of the 't Hooft renormalization scheme [10,11].  Such summation of logarithms within the effective potential has been considered in several models [18-23], as well as in a number of phenomenological [24,25] and effective action [26] applications.

The $\beta$-function  in $N = 2$ SYM theory is unaltered if the chiral superfield is given a mass (within a formulation of this model utilizing $N = 1$ superfields) [27-29]. If this mass is allowed to become large, the $N = 2$ SYM theory reduces to an $N = 1$ SYM theory as the chiral superfield decouples, leaving only the $N = 1$ real superfield. Indeed this is the same sort of decoupling that occurs in grand unified models [30,31]. According to the Appelquist-Carazzone theorm [32-35], the effect of such heavy fields is to renormalize the fields and couplings in an effective theory whose degrees of freedom are only those light fields present in the original theory. Consequently, it is possible to relate the $\beta$-function in $N = 2$ and $N = 1$ SYM theories. From this relationship we can deduce how the exact $\beta$-function in the SUSY renormalization scheme for $N = 1$ SYM theory is related to the MS $\beta$-function in the same model. The relationship between $N = 1$ and $N = 2$ SYM models based upon this decoupling has also been considered in ref. [36].

\section{The Effective Action in SYM Models}

The effective Lagrangian $\it{L}$ and the $\beta$-function have been argued [37] to be related by the equation
$$\textit{L} = -\frac{1}{4} \frac{g^2}{\overline{g}^2(t,g)} \Phi\eqno(1)$$
where
$$t = \frac{1}{4} \ln \left(\frac{g^2\Phi}{\mu^4}\right)\eqno(2)$$
and
$$\Phi = F_{\mu\nu}^a F^{a\mu\nu}.\eqno(3)$$
Eq. (1) is obtained by using the trace anomaly condition [13-15]
$$\left\langle \theta^\mu_\mu\right\rangle = \frac{\beta(\overline{g})}{2\overline{g}}\,
\left(\frac{g}{\overline{g}}\right)^2 \Phi\eqno(4)$$
in conjunction with the definition of the expectation value of the energy-momentum tensor
$$\left\langle \theta^{\mu\nu}\right\rangle = -g^{\mu\nu}\textit{L} + 2 \frac{\delta\textit{L}}{\delta g_{\mu\nu}}.\eqno(5)$$
To show that $\textit{L}$ of eq. (1) satisfies the renormalization group (RG) equation
$$\left(\mu \frac{\partial}{\partial\mu} + \beta (g) \frac{\partial}{\partial g} + \gamma (g) F_{\alpha\beta}^a \frac{\partial}{\partial F_{\alpha\beta}^a} \right) \textit{L} = 0\eqno(6)$$
we use the fact that [37,38]
$$\beta (g) = -g \gamma (g),\eqno(7)$$
which follows from the non-renormalization of $gA_\mu^a$ resulting from gauge invariance in the background field.

The function $\overline{g}(t,g)$ appearing in eq. (1) is defined by the equation
$$t = \int_g^{\overline{g}(t,g)} \frac{dx}{\beta(x)}\eqno(8)$$
so that
$$\frac{\partial\, \overline{g}(t,g)}{\partial t} = \beta(\overline{g} (t,g))\eqno(9)$$
$$\frac{\partial\, \overline{g}(t,g)}{\partial g} = \frac{\beta(\overline{g} (t,g))}{\beta (g)}\eqno(10)$$
and
$$\frac{\partial t}{\partial \mu} = -1.\eqno(11)$$

In ref. [16], eq. (1) was exploited to determine $\beta(g)$ to two-loop order by computing $\textit{L}$ to two-loop order with a self-dual background field. Here we proceed in the opposite direction; the exact $\beta$-function known for SYM theories is used to determine $\textit{L}$ via eq. (1) for the case of a background gauge field. 

Of course such a background necessarily breaks supersymmetry, as there is no Fermionic background field. In ref. [46] (as is discussed in ref. [48]) this sort of effective Lagrangian is discussed in $N = 4$ SYM theory using the Euler Heisenberg effective Lagrangian. A similar discussion occurs in ref. [47].

For $N = 1$ $SU(3)$ SYM theory, when using a SUSY renormalization scheme, the exact $\beta$-function is extracted from supersymmetry by [5-7] (see also ref. [51])
$$\beta (g) = \frac{-9 g^3/(4\pi)^2}{1 - 6 g^2/(4\pi)^2}\,.\eqno(12)$$
Integration of eq. (8) using the $\beta$-function of eq. (12) yields
$$\overline{y} e^{\overline{y}} = e^{-3t} ye^y\eqno(13)$$
where $\overline{y} = -(4\pi)^2/6\overline{g}^2 (t,g)$, $y = -(4\pi)^2/6g^2$. The Lambert $W$ function [39] is defined by 
$$W (\eta) e^{W(\eta)} = \eta\eqno(14)$$
and so eq. (1) becomes
$$\textit{L} = \frac{-1}{4y} W\left[\left(\frac{g^2\Phi}{\mu^4}\right)^{-3/4} ye^y\right]\Phi .\eqno(15)$$
From eq. (14), it follows that
$$W(\eta) = \sum_{n=1}^\infty \frac{(-n)^{n-1}}{n!} \eta^n\eqno(16)$$
and so eq. (15) becomes
$$\textit{L} = -\frac{1}{4} \left(\frac{-6g^2}{(4\pi)^2}\right)
\sum_{n=1}^\infty \frac{(-n)^{n-1}}{n!}\left[\left(\frac{g^2\Phi}{\mu^4}\right)^{-3/4}
\left(\frac{-4\pi^2}{6g^2}\right) \exp \left(\frac{-(4\pi)^2}{6g^2} \right)\right]^n\Phi\eqno(17)$$
which appears to be non-analytic at $g = 0$.

This action has a non-trivial extremum. To see this, note from eq. (14) that
$$W^\prime (\eta) = \frac{W(\eta)}{\eta(1 + W (\eta))}.\eqno(18)$$
The effective action of eq. (15) is of the form 
$$\textit{L}= A \cdot W \left[B\left( \frac{\Phi}{\mu^4}\right)^{-3/4} \right]\Phi .\eqno(19)$$
Then, we find that
$$\frac{1}{A}\, \frac{d\textit{L}}{d\Phi} = \left(\frac{1 + 4W}{1 + W}\right)W.\eqno(20)$$
Hence, if $\frac{dL}{d\Phi} = 0$, then either $W = 0$ (corresponding to $\Phi \rightarrow \infty$)or 
$$W\left[y\, e^y\left(\frac{g^2\Phi}{\mu^4}\right)^{-3/4}\right] = -\frac{1}{4}\, .\eqno(21)$$
If eq. (21) is satisfied, then the extremum\footnote{If this extremum is identified with $t = 0$ (hence $\Phi = \mu^4/g^2$) then this choice for the renormalization scale $\mu^2$ necessarily implies a non-perturbatively large value for the coupling ($g^2 = \frac{32\pi^2}{3}$).} of $\textit{L}$ occurs at
$$F_{\mu\nu}^a F^{a\mu\nu} = \frac{\mu^4}{g^2} \left[\frac{3}{2} \,\frac{g^2}{(4\pi)^2} \exp \left(- \frac{1}{4} + \frac{(4\pi)^2}{6g^2}\right)\right]^{-4/3}.\eqno(22)$$
Since by eq. (15), $L \rightarrow -\infty$, as $\Phi \rightarrow \infty$,  the extremum of eq. (21) is a maximum.

Similar behaviour arises with the mathematically simpler case of an $N = 2$ SYM theory, the $\beta$-function is given entirely by the one-loop expression
$$\beta(g) = -bg^3\eqno(23)$$
when employing the MS substraction scheme. For the gauge group $SU(3)$,
$$b = 6/(4\pi)^2 .\eqno(24)$$
Integration of eq. (8) yields
$$t = \frac{1}{2b} \left[ \frac{1}{\overline{g}^2(t,g)} - \frac{1}{g^2}\right]\eqno(25)$$
and hence from eq. (1)
$$\textit{L} = -\frac{1}{4} \left[1 + \left(\frac{g^2}{2}\right) b\ln \left(\frac{g^2\Phi}{\mu^4}\right)\right]\Phi .\eqno(26)$$
From eq. (26), if $\frac{dL}{d\Phi} = 0$, one obtains a non-zero extremum at
$$\Phi = \frac{\mu^4}{g^2} \exp \left(-1-\frac{2}{g^2b}\right)\eqno(27)$$
at which point the effective Lagrangian exhibits a maximum
$$\textit{L} = \frac{1}{8} b\mu^4 e^{-1-\frac{2}{g^2b}}.\eqno(28)$$
It is intriguing to speculate as to whether this maximum of the effective Lagrangian corresponds to a lower bound on a suitably defined effective potential for these theories. This is the point of view taken in ref. [40] where the one loop effective action in QCD is analyzed.

We now turn to relating the bare and renormalized couplings, using eqs. (12) and (23).

\section{Relating the Bare and Renormalized Couplings}

In ref. [17] the bare and renormalized couplings are considered for any model with a single coupling that is renormalized by minimal subtraction. It was shown there that the one-loop $\beta$-function can be used to sum all leading pole contributions to the bare coupling, the two-loop $\beta$-functions fixes the sum of the next-to-leading poles, etc. In the limit that we pass to four dimensions, each of these sums, and consequently the bare coupling, can be shown to vanish.

In general, the bare coupling $g_B$ and the renormalized coupling $g$ are related by
$$g_B = \mu^\epsilon\sum_{\nu = 0}^\infty \,\frac{a_\nu (g)}{\epsilon^\nu}\eqno(29)$$
where $\epsilon = 2 - d/2$ when working in $d$ dimensions and $\mu$ is a scale parameter introduced in the course of renormalization [41]. If we are using the MS renormalization scheme, then
$$a_0 (g) = g.\eqno(30)$$
We begin by considering the expression (29) when the minimal subtraction condition of eq. (30) is dropped. We now have
$$\mu \frac{dg_B}{d\mu} = 0 = \left( \mu \frac{\partial}{\partial \mu} + \mu \frac{dg}{d\mu}\,\frac{\partial}{\partial g}\right)g_B (\mu , g).
\eqno(31)$$
The cancellation of $O(\epsilon)$ contributions in eq. (31) implies that
$$\mu \frac{dg}{d\mu} = -\frac{a_0(g)}{a_0^\prime (g)} \epsilon + \beta (g)\eqno(32)$$
where $\beta(g)$ is the usual $\beta$-function in the $\epsilon \rightarrow 0$ limit. The cancellation of $O (\epsilon^0)$ terms in eq. (31) implies that
$$a_1(g) - \frac{a_0(g)}{a_0^\prime(g)} a_1^\prime (g) + \beta (g) a_0^\prime (g) = 0 \eqno(33)$$
in which case
$$\beta (g) = \frac{a_0^2(g)}{a_0^{\prime 2}(g)} \,\frac{d}{dg}\left(\frac{a_1(g)}{a_0(g)}\right) .\eqno(34)$$
Once the $\beta$-function is known through full knowledge of $a_0(g)$ and $a_1(g)$, the subsequent $a_k(g)$ are determined by the $O(\epsilon^{-k+1})$ contribution to eq. (31).
$$- \frac{a_0^2}{a_0^\prime} \,\frac{d}{dg} \left( \frac{a_k}{a_0}\right) + \beta a_{k-1}^\prime = 0\eqno(35)$$
so that
$$a_{k+1} (g) = a_0 (g) \int_0^g \frac{\beta (\lambda) a_0^\prime (\lambda) a_k^\prime (\lambda)}{a_0^2 (\lambda)}\,d\lambda . \eqno(36)$$

If now we define
$$\tilde{g} = a_0 (g)\eqno(37)$$
so that by eq. (30), $\tilde{g}$ is the MS renormalized coupling, then we find that
$$\mu \frac{d\tilde{g}}{d\mu} = -\epsilon \tilde{g} + \tilde{\beta}(\tilde{g}) = \frac{d\tilde{g}}{dg} \left(\mu \frac{dg}{d\mu}\right) = a_0^\prime (g) \left[ - \frac{a_0(g)}{a_0^\prime (g)} \epsilon + \beta(g)\right]\eqno(38)$$
$$\tilde{\beta}(\tilde{g}) = \tilde{\beta} (a_0(g)) = a_0^\prime (g) \beta(g)\eqno(39)$$
where $\tilde{\beta} (\tilde{g})$ is the $\beta$-function when using MS.

The form of the functions $a_\nu (g)$ is given by
$$a_\nu(g) = \sum_{n=\nu}^\infty a_{n,\nu} g^{2n+1}\eqno(40)$$
with
$$a_{0,0} = 1.\eqno(41)$$
The series in eq. (29) can now be reorganized so that
$$g_B = \mu^\epsilon g \sum_{k = 0}^\infty g^{2k} S_k \left(\frac{g^2}{\epsilon}\right)\eqno(42)$$
with
$$S_k (u) = \sum_{\ell = 0}^\infty a_{k+\ell ,k}u^\ell\,.\eqno(43)$$
We now see that eq. (31) and eq. (42) together imply that
$$\sum_{n=0}^\infty g^{2n} \left[\frac{g^3}{u} S_n(u) + \left(- \frac{a_0}{a_0^\prime} \frac{g^2}{u} + \beta(g)\right)\left((2n+1) S_n(u) + 2u S_n^\prime (u)\right)\right] = 0 .\eqno(44)$$

Furthermore, upon multiplying eq. (35)  by $\epsilon^{-k+1}$ and summing over $k$, we obtain
$$\epsilon\left(g_B - \frac{a_0(g)}{a_0^\prime (g)}\, \frac{\partial g_B}{\partial g}\right) + \beta (g) \frac{\partial g_B}{\partial g} = 0 ,\eqno(45)$$
a separable equation whose solution is
$$g_B = \exp \left[ - \int^g dx\, \frac{\epsilon}{\beta (x) - \epsilon \frac{a_0 (x)}{a_0^\prime (x)}} + K\right]\eqno(46)$$
which can be recursively employed to obtain summation over poles at $\epsilon = 0$ to all orders, as in ref. [17].
The constant $K$ in eq. (46) is fixed by requiring that eq. (41) is satisfied; we find that
$$g_B = \mu^\epsilon g \exp \left[ - \int_0^g 
\left(\frac{\epsilon}{\beta (x) - \epsilon \frac{a_0 (x)}{a_0^\prime (x)}} + \frac{1}{x}\right) dx \right].\eqno(47)$$
(This expression is close to eq. (7.5) of ref. [41].)

Formally, $g_B$ is independent of $\mu$ but not of the dimensionality parameter $\epsilon$. We see immediately that irrespective of $a_0$ (which is not known in $N = 2$ SYM theory even if $\beta$ is given exactly by eq. (12)), eq. (47) involves an improper integral in the four dimensional ($\epsilon \rightarrow 0$) limit
$$\lim_{\epsilon \rightarrow 0} g_B = g\exp \left(-\lim_{\delta \rightarrow 0^+} \int_\delta^g \frac{dx}{x} \right) = 0.\eqno(48)$$
This explicit vanishing of the bare coupling in four dimensions is noted in the MS scheme by more elaborate means in ref. [17]. However, the above result pertains to all schemes.

As an explicit example, consider eq. (47) for the $N = 2$ SYM $\beta$-function of eq. (23):
$$g_B = \mu^\epsilon g\exp \left[-\int_0^g \left(\frac{\epsilon}{-bx^3 - \epsilon x} + \frac{1}{x}\right)\right]\nonumber$$
$$ = \mu^\epsilon g\left(1 + \frac{b}{\epsilon} g^2 \right)^{-1/2}.\eqno(49)$$
This closed form expression clearly shows that $g_B$ vanishes in the limit $\epsilon \rightarrow 0$. Eq. (49) can also be obtained by interating eq. (36) and then performing the sum of eq. (29) explicitly.

\section{The Effective Potential in a $\phi_4^4$ Model}

We now examine the effective potential in a massless $\phi_4^4$ model, as introduced in [42-45]. The general form of this potential is
$$V(\phi , \lambda, \mu) = \lambda \phi^4 \sum_{n=0}^\infty \sum_{m=0}^n a_{n,m} \lambda^n L^m \eqno(50)$$
where $L = \ln (\phi/\mu)$. The double sum of eq. (50) can be reorganized into a sum of ``leading logarithms'' (LL), ``next-to-leading logarithms'' (NLL) etc., 
$$V(\phi, \lambda, \mu) =  \sum_{n=0}^\infty \lambda^{n+1} S_n(\lambda L) \phi^4 \eqno(51)$$
where $n$ is the index characterizing the summation of $N^nLL$ ($n = 0$ is $LL$):
$$S_n(\xi) = \sum_{m=0}^\infty a_{m+n,m}\xi^m\,.\eqno(52)$$
For $V$ to be independent of the scale parameter $\mu$, then the RG equation must be
satisfied
$$\mu \frac{dV}{d\mu} = 0 = \left(\mu \frac{\partial}{\partial \mu} + \beta (\lambda) \frac{\partial}{\partial \lambda} + \gamma (\lambda) \phi\frac{\partial}{\partial\phi}\right) V(\phi , \lambda , \mu)\;;\eqno(53)$$
where $\beta(\lambda) = \mu \frac{d\lambda}{d\mu}$ and $\gamma(\lambda) = \frac{\mu}{\phi}\,\frac{d\phi}{d\mu}$. It is possible to work within the context of the 't Hooft renormalization scheme [10,11] where $\beta(\lambda)$ is given exactly by the two-loop result
$$\beta(\lambda) = b_2\lambda^2 + b_3\lambda^3\eqno(54)$$
and $\gamma(\lambda)$ by the one-loop result
$$\gamma(\lambda) = c\lambda^2 \,.\eqno(55)$$
It is known from explicit calculation that $b_2 = \frac{3}{16\pi^2}$, $b_3 = \frac{-17}{3(16\pi^2)^2}$ and $c = 0$. Substitution of eqs. (51), (54) and (55) into eq. (53) leads to a succession of differential eqs. for $S_n$:
$$(-1 + b_2\xi) \frac{dS_0(\xi)}{d\xi} + b_2 S_0 (\xi) = 0\eqno(56)$$
$$\left(-1 + b_2\xi\right) \frac{dS_n(\xi)}{d\xi} + b_2 (n+1)S_n(\xi) + b_3 \left(\xi \frac{d}{d\xi} + n\right) S_{n-1}(\xi) = 0.\eqno(57)$$
If derivatives are now defined with respect to $w = 1 - b_2\xi$, and $r = b_3/b_2$, then eq. (57) becomes
$$\left(S^\prime_n + \frac{n+1}{w} S_n\right) + r \left(\frac{w-1}{w} S^\prime_{n-1} + \frac{n}{w} S_{n-1}\right) = 0; \eqno(58)$$
integrating eq. (58) gives
$$S_n(w) = w^{-n-1}\left[-r\int_1^w dt\,t^n\left( (t-1) S_{n-1}^\prime (t) + n S_{n-1}(t)\right) + a_{n,0}\right]\eqno(59)$$
with $a_{0,0} =1 $. Since
$$(t-1)S_{n-1}^\prime (t) + n S_{n-1} (t) = (t-1)^{1-n} \frac{d}{dt} \left((t-1)^{n} S_{n-1}(t)\right),\eqno(60)$$
eq. (59) can be integrated by parts to yield
$$T_n(w) = -\left[(w-1) T_{n-1}(w) + \int_1^w \left(\frac{n}{t} - 1\right) T_{n-1} (t) dt + \frac{a_{n,0}}{r^n}
\right]\eqno(61)$$
where
$$S_n(w) = r^n T_n(w)/w^{n+1}\eqno(62)$$
and $T_0(w) = 1$. By iterating eq. (61), $V$ can be determined in terms of $a_{n,0} (n = 0, 1, 2 \ldots)$. These quantities characterize the 't Hooft renormalization scheme where eqs. (54) and (55) are exact.

From eq. (61) we find that
$$T^\prime_{n+1}(w) = (1-w) T^\prime_n(w) - \left(  \frac{n+1}{w}\right)T_n(w)\eqno(63)$$
which implies that the form of $T_n(w)$ is given by
$$T_n (w) = \sum_{k=0}^n \sum_{\ell = 0}^{n-k} \kappa_{k,\ell}^{(n)} w^\ell (\ln w)^k\eqno(64)$$
with 
$$\kappa_{0,0}^{(n)} = a_{n,0}.\eqno(65)$$
Upon substitution of eq. (64) into eq. (63), we get
$$\ell\, \kappa^{(n+1)}_{k-1,\ell} + k\kappa^{(n+1)}_{k,\ell} = (\ell - n - 1)\kappa_{k-1,\ell}^{(n)} - (\ell - 1) \kappa_{k-1,\ell -1}^{(n)}\eqno(66)$$
$$+ k\left(\kappa_{k,\ell}^{(n)} - \kappa_{k,\ell -1}^{(n)}\right)\nonumber$$
where $\kappa_{k,\ell}^{(n)}$ is zero if $k,\ell$ do not lie within the range $0 \leq k \leq n$; $0 \leq \ell \leq n-k$.

Although finding a closed form expression for $\kappa_{k,\ell}^{(n)}$ is prohibitively difficult, one can determine $\kappa_{n,0}^{(n)}$ and $\kappa_{0,n}^{(n)}$ by setting $\ell = 0, k = n + 1$ and $k = 1, \ell = n + 1$ respectively in eq. (66).

We easily find that
$$\kappa_{n,0}^{(n)} = -\kappa_{n+1, 0}^{(n+1)} = (-1)^n\eqno(67)$$
and similarly
$$\kappa_{0,n}^{(n)} = 0.\eqno(68)$$
The contribution to $V$ coming from these two sets of coefficients is by eq. (51)
$$\tilde{V} (\phi , \lambda , \mu) = \frac{\lambda\phi^4}{w} \sum_{n=0}^\infty \left(\frac{-\lambda r\ln w}{w}\right)^n
\nonumber$$
$$= \lambda\phi^4\left(\frac{1}{w + \lambda r\ln w}\right) .\eqno(69)$$
This all-orders contribution to $V(\phi , \lambda , \mu)$ develops a peculiar singularity $w + \lambda r \ln w = 0$.

\section{Relating Renormalization Schemes in $N = 2$ SYM}

The $N = 2$ SYM model can be viewed as a real  $N = 1$ SYM vector superfield coupled to a massless chiral superfield, provided both are in the adjoint representation of the gauge group. If the chiral superfield is given a mass, then the ultraviolet properties of the model are not altered and the $\beta$-function is not changed. Upon letting this mass become much larger than any other mass scale in a physical process, the chiral superfield field decouples and what remains is an effective theory whose dynamics is that of the residual $N = 1$ SYM vector superfield. The Appelquist-Carazzone theorem [32-35] shows that to leading order the effect of this massive chiral superfield is to renormalize the parameters characterizing the effective theory which is an $N = 1$ SYM model. Quantitatively, this means that
$$\Gamma_n (p,g,M,\mu) = Z^n(g,M,\mu)\Gamma^*_n (p,g^*(g,M,\mu),\mu) + O \left(\frac{1}{M^2}\right).\eqno(70)$$
Here $\Gamma_n$ is the $n$-point Green's function calculated in the full $N = 2$ SYM model supplemented by its chiral superfield being given a mass $M$, while $\Gamma_n^*$ is the analogous Green's function in the effective $N = 1$ SYM theory that remains when $M^2$ is large. Its coupling is $g^*$.

These Green's functions both satisfy $RG$ equations
$$\left[\mu \frac{\partial}{\partial\mu} + \beta (g) \frac{\partial}{\partial g} + \gamma_M (g) M \frac{\partial}{\partial M} - n \gamma (g) \right] \Gamma_n (p, g, M, \mu) = 0\eqno(71)$$
and
$$\left[\mu \frac{\partial}{\partial\mu} + \beta^* (g^*) \frac{\partial}{\partial g*} - n  \gamma^* g^* \right]  \Gamma_n^* (p, g^*, \mu) = 0.\eqno(72)$$
Together, eqs. (70-72) imply that
$$\beta^*\left(g^*(g,M,\mu)\right) = \left(
\mu \frac{\partial}{\partial \mu} + \beta (g) \frac{\partial}{\partial g} +  \gamma_M (g) M \frac{\partial}{\partial M}\right)g^*(g, M, \mu).\eqno(73)$$
Eq. (70) holds only if both $\Gamma_n$ and $\Gamma_n^*$ are computed using the same renormalization scheme. If now the SUSY scheme $\beta$-function of eq. (12) has renormalized coupling $g^*$ while the $MS$ renormalized coupling is $\tilde{g}^*$, then
$$\tilde{g}^* = a_0(g^*)\eqno(74)$$
where $a_0(g^*)$ is the leading term in the expansion of eq. (29) when employing the SUSY renormalization scheme. We then can consider the relationship between $\tilde{g}$, the renormalized coupling in the $N = 2$ SYM model when employing the MS renormalization scheme, and $g$, defined by
$$\tilde{g} = a_0(g).\eqno(75)$$
If now
$$a_0 (g) = g + \alpha_3 g^3  + \alpha_5 g^5 + \alpha_7 g^7 + \ldots\nonumber$$
then by eqs. (39) and (23),
$$-b \left(g + \alpha_3 g^3 + \alpha_5 g^5 + \ldots \right)^3
= \left(-b_3 g^3 - b_5 g^5 - b_7 g^7 - \ldots\right)
\left(1 + 3\alpha_3 g^2 + 5\alpha_5 g^4 + \ldots \right)\eqno(76)$$
where
$$\beta (g) = -\sum_{k = 1}^\infty b_{2k+1} g^{2k+1} .\eqno(77)$$
From eq. (76) we find that
$$b_3 = b, \;\;b_5 = 0,\;\; b_7 = \left(3\alpha_3^2 - 2\alpha_5\right)b,\eqno(78)$$
etc.

Integration of the equation
$$\mu \frac{dg^*}{d\mu} = \beta^* (g^*) \eqno(79)$$
with $\beta^*(g^*)$ given by eq. (12) with the boundary condition $g^* = g$ when $\mu = M$ leads to
$$y^* e^{y*} = \left(\frac{\mu}{M}\right)^{-3} y^y\eqno(80)$$
where
$$y^* = \frac{-(4\pi)^2}{6g^{*2}}\eqno(81)$$
and
$$y = \frac{-(4\pi)^2}{6g^2} .\eqno(82)$$
(Eqs. (13) and (80) are solutions to the same equation; they arise in different contexts.) From eqs. (80-82) we see that 
$$g^{*2} = - \frac{(4\pi)^2}{6} \left[W \left(\left( \frac{\mu}{M}\right)^{-3} y^y\right)\right]^{-1}.\eqno(83)$$

We also know that in a supersymmetric field theory, the mass term in the action for a chiral superfield $\phi$, $M\phi^2|_F + h.c.$, is not renormalized and hence (using eq. (7))
$$\gamma_m (g) = -2\gamma(g) = 2\beta (g)/g\eqno(84)$$
regardless  of the renormalization scheme employed. We now can consider eq. (73) in the SUSY renormalization scheme. In this equation, by eqs. (12) and (83)
$$\beta^* (g^* (g, M, \mu)) = 
\frac{-\frac{9}{(4\pi)^2} \left[-\frac{(4\pi)^2}{6} W^{-1}\left(\left(\frac{\mu}{M}\right)^{-3}y^y\right)\right]^{3/2}}{1 + W^{-1}\left(\left(\frac{\mu}{M}\right)^{-3}y^y\right)}\eqno(85)$$
while by eqs. (39) and (23)
$$\beta (g) = \frac{1}{a_0^\prime (g)} \left[-b (a_0 (g))^3\right].\eqno(86)$$
Together, eqs. (84-86) result in
$$\frac{
\frac{-9}{(4\pi)^2}\left[-\frac{(4\pi)^2}{6} W^{-1}\left(\left(\frac{\mu}{M}\right)^{-3}y^y\right)\right]^{3/2}}
{1 + W^{-1}\left(\left(\frac{\mu}{M}\right)^{-3}y^y\right)}
= \left\lbrace\left[-1 + \frac{2}{a_0^\prime(g)} \left(-b (a_0(g))^2\right)\right]\frac{\partial}{\partial L}\right. \eqno(87)$$
$$\left. + \frac{1}{a_0^\prime(g)} \left[-b (a_0(g))^3 \right] \frac{\partial}{\partial g}\right\rbrace\left[- \frac{(4\pi)^2}{6} W^{-1}\left(\left(\frac{\mu}{M}\right)^{-3} y^y\right)\right]$$
where $L = \ln \left(\frac{M}{\mu}\right)$ and $y$ is given by eq. (82). Although eq. (87) is not particularly tractable, it does in principle provide a method of fixing the function $a_0 (g)$. This function need not be unique [7].

\section{Discussion}

In this paper, we have examined a number of instances in which exact knowledge of the perturbative $\beta$-function allows one to determine a number of physical results without recourse to explicit calculation. In particular, the effective action for a background $U(1)$ vector field is fixed in $N = 1$ and $N = 2$ SYM theory, and the effective potential in a $\phi_4^4$ model has all its logarithmic parts determined when in the 't Hooft renormalization scheme. The way in which bare and renormalized couplings are related in $N = 1$ and $N = 2$ SYM models is also fully determined. 
Finally the finite renormalization required to convert the SUSY coupling to the MS coupling in $N = 1$ SYM theory is examined.

\section{Acknowledgements}

We would like to thank A. Buchel, G. Dunn, V. Miransky and C. Schubert for discussions. R. and D. MacKenzie had numerous valuable suggestions. NSERC provided financial support. The hospitality of the Perimeter Institute for Theoretical Physics where this work was completed is appreciated.


\begin{thebibliography}{99}
\bibitem{1} S. Mandelstam, Nucl. Phys. B213, 149 (1983).
\bibitem{2} P. Howe, K. Stelle and P. Townsend, Nucl. Phys. B214, 519 (1983).
\bibitem{3} P. Howe, K.S. Stelle and P. West, Phys. Lett. 124B, 55 (1983).
\bibitem{4} I.G. Koh and S. Rajpoot, Phys. Lett. 135B, 397 (1983).
\bibitem{5} V. Novikov, V. Shifman, A. Vainshtein and V. Zakharov, Nucl. Phys. B229, 381 (1983).
\bibitem{6} D.R.T. Jones, Phys. Lett. 123B, 45 (1983).
\bibitem{7} V. Elias, J. Phys. G. 27, 217 (2001).
\bibitem{8} I. Jack, D.R.T. Jones and C.G. North, Nucl. Phys. B486, 479(1997).
\bibitem{9} I. Jack. D.R.T. Jones and A. Pickering, Phys. Lett. B435, 61 (1998).
\bibitem{10} G. 't Hooft, Erice Lectures 1977 (unpublished).
\bibitem{11} N.N. Khuri and O.A. McBryan, Phys. Rev. D20, 881 (1979).
\bibitem{12} R. Crewther, Phys. Rev. Lett. 28, 1421 (1972).
\bibitem{13} S.L. Adler, J.C. Collins and A. Duncan, Phys. Rev. D15, 1712 (1977).
\bibitem{14} J.C. Collins, A. Duncan and S.D. Joglekar, Phys. Rev. D16, 438 (1977).
\bibitem{15} N.K. Nielsen, Nucl. Phys. B120, 212 (1977).
\bibitem{16} G.V. Dunne, H. Gies and C. Schubert, JHEP 0211, 032 (2002).
\bibitem{17} V. Elias and D.G.C. McKeon, Int. J. Mod. Phys. A18, 2395 (2003).
\bibitem{18} B. Kastening, Phys. Lett. B 283, 287 (1982).
\bibitem{19} C. Ford, Phys. Rev. D50, 7531 (1994).
\bibitem{20} M. Bando, T. Kugo, M. Maskawa and H. Nakano, Phys. Lett. B301, 83 (1993).
\bibitem{21} J. M. Chung and B.K. Chung, Phys. Rev. D60, 105001 (1999).
\bibitem{22} V. Elias, R.B. Mann, D.G.C. McKeon and T.G. Steele, Phys. Rev. Lett. 91, 251601 (2003).
\bibitem{23} V. Elias, R.B. Mann, D.G.C. McKeon and T.G. Steele, Nucl. Phys. B678, 147 (2004).
\bibitem{24} M. Ahmady, F. Chishtie, V. Elias, A. Fariborz, N. Fattahi, D.G.C. McKeon, T.N. Sherry and T.G. Steele, Phys. Rev. D66, 014010 (2002).
\bibitem{25} D.G.C. McKeon and A. Rebhan, Phys. Rev. D67, 027701 (2003).
\bibitem{26} M. Ahmady, V. Elias, D.G.C. McKeon, A. Squires and T.G. Steele, Nucl. Phys. B655, 221 (2003).
\bibitem{27} S. Rajpoot, J.G. Taylor and M. Zaimi, Phys. Lett. 127B, 347 (1983).
\bibitem{28} A. Parkes and P. West, Nucl. Phys. B222, 269 (1983).
\bibitem{29} M.A. Namatic, A. Salam and J. Strathdee, Phys. Rev. D28, 1481 (1983).
\bibitem{30} S. Weinberg, Phys. Lett. 82B, 387 (1979).
\bibitem{31} L. Hall, Nucl. Phys. B178, 75 (1981).
\bibitem{32} T. Appelquist and J. Carazzone, Phys. Rev. D11, 2856 (1975).
\bibitem{33} E. Witten, Nucl. Phys. B104, 445 (1976).
\bibitem{34} F. Gilman and M.B. Wise, Phys. Rev. D20, 2392 (1979).
\bibitem{35} D.G.C. McKeon, Phys. Rev. D26, 2086 (1982).
\bibitem{36} D.G.C. McKeon and S. Rajpoot, Phys. Lett. 151B, 229 (1985).
\bibitem{37} S.G. Matinyan and G.V. Savvidy, Nucl. Phys. B134, 539 (1978).
\bibitem{38} L. Abbott, Nucl. Phys. B185, 189 (1981).
\bibitem{39} R.M. Corless, G.H. Gonnet, D.E.G. Hare, D.J. Jeffrey and D.E. Knuth, Adv. Comp. Math. 5, 329 (1996).
\bibitem{40} J. Ambj{\o}rn, N.K. Nielsen and P. Olesen, Nucl. Phys. B152, 75 (1979).
\bibitem{41} G. 't Hooft, Nucl. Phys. B61, 455 (1973).
\bibitem{42} S. Coleman and E. Weinberg, Phys. Rev. D7, 1888 (1973).
\bibitem{43} S. Weinberg, Phys. Rev. D7, 2887 (1973).
\bibitem{44} R. Jackiw, Phys. Rev. D9, 1686 (1974).
\bibitem{45} A. Salam and J. Strathdee, Phys. Rev. D9, 1129 (1974).
\bibitem{46} E.S. Fradkin and A. A. Tseytlin, Nucl. Phys. B227, 252 (1983).
\bibitem{47} D.G.C. McKeon, I. Sachs, I.A. Shovkovy, Phys. Rev. D59, 105010 (1999).
\bibitem{48} G. Dunne, hep-th 0406216.
\bibitem{49} G. Veneziano and S. Yankielowicz, Phys. Lett. 113B, 231 (1982).
\bibitem{50} T.R. Taylor, G. Veneziano and S. Yankielowicz, Nuc. Phys. B218, 493 (1983).
\bibitem{51} N. Arkani-Hamed and H. Murayama, JHEP 06, 030 (2000).
\end{thebibliography}
\end{document}